\documentstyle[12pt,epsf]{article}
\begin{document}
\rightline{CU-TP-983}
\begin{center}
{{\Large\bf On the Relationship between Large Order Graphs and Instantons
for the Double Well Oscillator}\footnote{This work is
supported in part by the U.S. Department of Energy}\\
\vskip 15pt
A.H. Mueller and D.N. Triantafyllopoulos\footnote{e-mail address: 
dionysis@phys.columbia.edu}\\ Department of Physics, Columbia University\\
New York, New York 10027}
\end{center}

\vskip 20pt
\centerline{\bf Abstract}

The double well oscillator is used as a QCD-like model for
studying the relationship between large order graphs and the
instanton-antiinstanton solution.  We derive an equation for
the perturbative coefficients of the ground state energy
when the number of 3 and/or 4-vertices is fixed and large. 
These coefficients are determined in terms of an exact
``bounce" solution.  When the number of 4-vertices is
analytically continued to be near the negative of half the
number of 3-vertices the bounce solution approaches the
instanton-antiinstanton solution and determines leading Borel
singularity.

\section{Introduction}

\indent The QCD perturbation series diverges in a factorial
manner\cite{Eri}.  If A is the amplitude for an infrared safe
process and if $A_n\alpha^n$ is the $n^{th}$ order term in
its perturbative expansion one expects terms of the form
$A_n\sim cn!R^n n^\gamma$ at large orders.  While $\gamma$
depends on the amplitude in question, $R$ is independent of
$A.$  The only known sources for such factorial terms are
infrared renormalons, ultraviolet renormalons
and instantons\cite{Eri,Ben}.  At large orders the logarithmic
dependence of the running coupling gives $n!$ terms.  When
these terms come from low momentum regions they give
singularities (infrared renormalons) in the right-half plane
of the Borel function corresponding to A.  Such singularities
signal a breakdown of the perturbative expansion and
correspond to higher twist terms, requiring nonperturbative
information, when that expansion is done for a hard process. 
Running coupling effects in the ultraviolet regime lead to
singularities in the left-half Borel plane which are
harmless and do not obstruct the Borel resummation of the
series.  The third type of singularities, nonsummable
singularities in the right-half Borel plane, can be viewed in 
two seemingly different ways.  On the one hand the leading
singularity of this variety is believed to correspond to a
well separated instanton-antiinstanton pair having action
$4\pi/\alpha$ and leading to a singularity at $b=4\pi$ in the
Borel plane.  On the other hand,  this singularity is
believed to be due to graph counting and as such should be
visible in large order perturbative calculations.  The
precise relationship between the Feynman graph  picture and
the instanton picture has never been given and it is this
task which motivates the present work.

In order to gain an understanding of what the relationship
between large order graphs and instantons might be, we study
here the analogous question for the quantum mechanical double
well oscillator.  Indeed it is well-known that there is a
simple relationship between the instantons in the double
well oscillator and the QCD instantons.  Since graph counting
only depends on the types of vertices in the theory, but not
on whether a field theory or a quantum mechanical problem is
being considered, one can expect this aspect of the problem
also to be similar for the double well oscillator and for QCD.

\indent The relationship between classical solutions and graphs is
understood for the anharmonic oscillator\cite{Lip,Par,Zub,Itz}, 
$S=\int_{-\infty}^\infty dt[{1\over 2}\dot{\varphi}^2(t)+{1\over 2}
\varphi^2(t)+{1\over 2}\alpha \varphi^4(t)]$ in Euclidean time. For large $n$ 
the evaluation of
the
$n^{th}$  order coefficient, $A_n,$ in a perturbative series in $\alpha$ is 
determined by the classical solution to the Lagrangian after making the 
replacement $\alpha \to -2/3n.$  One may view $A_n$ 
as being given by the number of graphs at order  \ $n$\  times the
value of a typical graph\cite{Par}, which value is determined by a
classical (bounce) solution.

\indent We follow the same procedure in the double well oscillator whose
(Euclidean) Lagrangian is given in (1).  Since we are interested in
evaluating large order graphs we find it useful to write the
perturbative expansion in terms of the number of 3-vertices, $2n_3,$
and the number of 4-vertices, $n_4,$ as indicated in (26). 
Following the Lipatov\cite{Lip,Itz} procedure we are able in Sec.2 to
evaluate $E(n_3,n_4),$ the sum of all graphs having $2n_3$ 3-vertices
and $n_4$ 4-vertices, in terms of a classical solution to (12) and
(16).  However, this classical solution seems to be unrelated to the
instanton solution of the original double well potential.  This
classical solution (20) and (21) and the evaluation of the small oscillations
about it are given in Sec.3.  The evaluation of  the coefficient of
$\alpha^n, n=n_3 + n_4,$ involves the sum $\sum_{n_4=0}^nE(n-n_4,n_4)
= -\sum_{n_4=0}^n(-1)^{n_4}\vert\ E(n-n_4,n_4)\vert,$ an alternating
sign series with delicate cancellations.  Nevertheless, the classical
solution can be used to compute $E(n_3,n_4)$ with sufficient accuracy
to do the sum and generate the perturbation series in $\alpha^n.$  This
is done in Sec.4 where all the $E(n_3,n_4)$ are given (numerically) for
$n_3+n_4=10.$

\indent However, we still have not made a connection between graphs and the
instanton-antiinstanton solution of the double well potential.  To do
this we go to the Borel plane and look at the singularity at $b=1/3$
which is known to be related to the instanton-antiinstanton
solution\cite{Bre,Zub,Bog,Bal,Fal}, and which we are going to
evaluate in terms of  graphs.  The integrand of the Borel
representation is given in Sec.5.  In Secs. 6 and 7 we evaluate the
Borel integrand and find the singularity at $b=1/3.$  In this
evaluation we face the same alternating sign series involving
$E(n_3,n_4) = -\vert E(n_3,n_4)\vert (-1)^{n_4}.$  To avoid this
alternating sign we write the $n_4$ sum as a Sommerfeld-Watson integral first in $n_4$ 
and later
in
$r = - n_4/n_3.$  For $b$ near $1/3$ the pole part of the Borel integrand is
determined by values of $r$ and $n_3$ given by $n_3 \propto [{1\over 3}
- b]^{-1}\vert log (1/3-b) \vert$ and $1-r \propto \vert log (1/3-b)\vert^{-1}.$  Thus the
instanton-antiinstanton solution corresponds to a graph with a large number, $2n_3,$ of 3-vertices
and a large \underline{negative} number, $-rn_3,$ of 4-vertices with the
instanton-antiinstanton separation given by $T={2\over 3}(1-r)^{-1}.$ 
We have no simple interpretation of why the instanton-antiinstanton
pair corresponds to an (analytically continued) negative number of
4-vertices.

\indent In Sec.8, we discuss the instanton-antiinstanton solution as the
analytic continuation of the bounce solution we found in Sec.3 and
comment on the zero mode.  We emphasize, however, that (55) is an
\underline{exact} solution to (12) and closely resembles the usual
$I-\bar{I}$ pair when $\epsilon$ is small.  We note that we have no
quasi-zero mode\cite{Yun} in our formalism because the
instanton-antiinstanton separation is fixed in terms of the value of
$1/3-b$ in the Borel integrand.

In Appendix A we evaluate integrals which appear while in Appendix B
we evaluate the determinant of the small fluctuations about the
classical solution describing large order graphs.

\section{Saddle point of the effective action and large orders of the perturbation series}

\indent We start from the Euclidean time Lagrangian for a double well
anharmonic oscillator with degenerate vacua at $\varphi = 0$ and
$\varphi =1/{\sqrt{\alpha}}.$  The potential is shown in Fig.1 and the
Lagrangian is given by

\begin{equation}
L={1\over 2}{\dot\varphi}^2 + {1\over 2} \varphi^2- {\sqrt{\alpha}}\
\varphi^3 + {1\over 2} \alpha \varphi^4.
\end{equation}

\noindent We want to calculate the large order in $\alpha$ behavior of
the ground state energy $E(\alpha)$ which is given by the path integral

\begin{equation}
E(\alpha) - E(0) = \lim_{\beta \to \infty}\left(-{1\over \beta} log
Z(\alpha)\right)
\end{equation}

\noindent where

\begin{equation}
Z(\alpha) =  {1\over Z(0)} \int D\varphi e^{-S[\varphi]}
\end{equation}

\begin{equation}
Z(0) = \int D\varphi e^{-S_0[\varphi]},
\end{equation}

\noindent with 
\begin{equation}
S[\varphi] = {1\over 2} \int dt({\dot\varphi}^2(t) +
\varphi^2(t)) -{\sqrt{\alpha}}\int dt \varphi^3(t) + {1\over 2} \alpha
\int dt \varphi^4(t).
\end{equation}

\noindent $S_0[\varphi]$ is the action of the {\em{free}} theory defined by

\begin{equation}
L_0={1\over 2} {\dot\varphi}^2 + {1\over 2}\  \varphi^2.
\end{equation}

\begin{figure}[htb]
\begin{center}
\epsfbox[0 0 211 135]{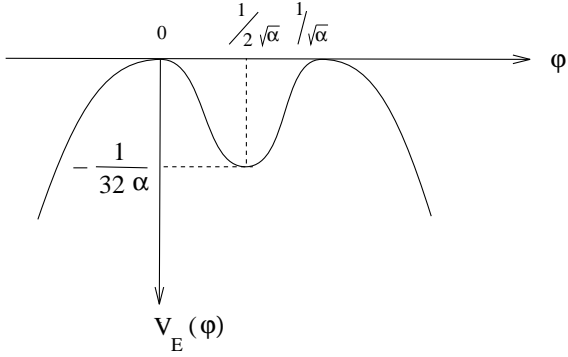}
\caption{The potential in Euclidean time.}
\end{center}
\end{figure}

\noindent In the path integral in (3) the configurations $\varphi(t)$
satisfy the {\em{periodic}} boundary condition

\begin{equation}
\varphi(-\infty) = \varphi(\infty).
\end{equation}

\indent To construct the perturbation series one expands the
exponentials of the interaction terms of  $S$  in (3) to obtain

\begin{equation}
Z(\alpha)\ =\ {1\over Z(0)} \sum_{n_3,n_4}\
{\alpha^{n_3+n_4}(-1)^{n_4}\over (2n_3)!n_4!2^{n_4}}\int
D\varphi\ e^{-S_{eff}[\varphi]}
\end{equation}

\noindent where the effective action is

\begin{equation}
S_{eff} =\int dt({1\over 2}{\dot\varphi}^2+{1\over 2}\varphi^2) -2n_3 log \int
dt \varphi^3-n_4 log \int dt \varphi^4.
\end{equation}

\noindent Next, in order to evaluate the terms in the sum in (8) when
$n_3$ and/or $n_4$ are large we look for a saddle point $\varphi_c(t),$
a classical solution, which minimizes the effective action.  Then we use
the steepest descent method to evaluate the path integral in (8) in
terms of paths near the classical solution. Thus we can write

\begin{equation}
Z(\alpha) = \sum_{n_3,n_4}\ {\alpha^{n_3+n_4}
2(-1)^{n_4}{\cal{N}}_\varphi e^{-S_{eff}[\varphi_c]}\over (2n_3)! n_4!
2^{n_4}}.
\end{equation}

\noindent$\cal{N}_\varphi$ is the prefactor coming from the integration
around $\varphi_c$ divided by the normalization factor $Z(0).$  This
small fluctuation integration is Gaussiasn after the zero mode is
isolated.  The zero mode is a consequence of the translational
invariance of the problem.  If $\varphi_c(t)$ is a solution, then
$\varphi_c(t-\tau)$ is also a solution for any $\tau.$  The factor of
2 in (10) appears because there is a discrete symmetry.  If $\varphi_c$
is a solution, then $-\varphi_c$ is also one.  Now the solution is
found from

\begin{equation}
{\delta S_{eff}\over \delta \varphi(t)}{\Big{\vert}}_{\varphi_c}=0,
\end{equation}

\noindent which leads to

\begin{equation}
K_0 \varphi_c = {6n_3\over \int dt \varphi_c^3} \varphi_c^2 +
{4n_4\over \int dt \varphi_c^4} \varphi_c^3
\end{equation}

\noindent where

\begin{equation}
K_0= {-d^2\over dt^2} + 1.
\end{equation}

\indent In order to simplify the differential equation rescale the
positive $\varphi_c$ according to

\begin{equation}
\varphi_c = {\sqrt{{2n_3\over J_3}}}  {\tilde\varphi}_c,
\end{equation}

\noindent and define the integrals

\begin{displaymath}
J_2 = \int dt {\tilde\varphi}_c^2, \ \ \ \ \ \ \ \ \  \dot{J}_2 = \int dt
\dot{\tilde\varphi}_c^2
\end{displaymath}

\begin{equation}
J_3 = \int dt {\tilde\varphi}_c^3,\ \ \ \ \ \ \ \ \   J_4
= \int dt {\tilde\varphi}_c^4.
\end{equation} 

\noindent Then Eq.(12) becomes

\begin{equation}
K_0{\tilde\varphi}_c = 3{\tilde\varphi}_c^2 + 2(\epsilon^2-1) {\tilde\varphi}_c^3
\end{equation}

\noindent where, for convenience, we have defined the parameter

\begin{equation}
\epsilon^2=1 + {n_4\over n_3}\ {J_3\over J_4}.
\end{equation}

\noindent Notice that $\epsilon^2$ is the only parameter appearing in Eq.(16).  The solution
${\tilde\varphi}_c$ and therefore the integrals $J_3, J_4$ depend only on this single parameter. 
Thus, given the ratio $n_4/n_3, \epsilon^2$ can be found from (17).  In other words every quantity
which is a function of $\epsilon^2$ can be considered a function of $n_4/n_3.$  For fixed
$\epsilon^2,$ one can give an effective potential leading to (16).  This potential is stable at
infinity since $\epsilon^2 > 1$ as can be seen from the defining equation (17).  The effective
potential is shown in Fig.2 where the dotted line shows the classical path
$\tilde{\varphi}_c$ which is positive.  Before moving on to find
$\varphi_c,$ let's calculate $S_{eff}[\varphi_c]$ in terms of $\epsilon^2,n_3,n_4.$  Multiplying
(16) by ${\tilde\varphi_c}$ and integrating one obtains

\begin{equation}
J_2+\dot{J}_2 =\left(3 + {2n_4\over n_3}\right) J_3
\end{equation}

\noindent where we have used (17).  Substitution of (14) and (18) into (9) gives

\begin{equation}
S_{eff}[\varphi_c]= 3n_3+2n_4-n_3 log {8n_3^3\over J_3} - n_4 log {4n_3^2 J_4\over J_3^2}.
\end{equation}

\begin{figure}
\begin{center}
\epsfbox[0 0 166 103]{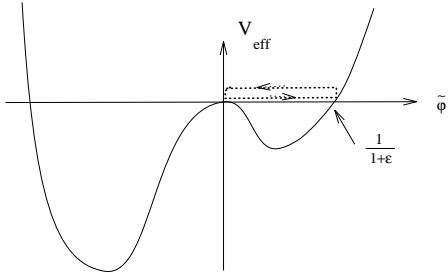}
\caption{The ``effective'' potential for fixed $\epsilon^2>1.$  The dotted line represents the
bounce solution $\tilde{\varphi}_c.$}
\end{center}
\end{figure}

\section{The classical solution}

Given the ratio $n_4/n_3$ the solution of equation (16) is unique when one imposes the boundary
condition, but there is a freedom to make time translations.  Integrating twice, the positive
solution that satisfies (16) is

\begin{equation}
\tilde{\varphi}_c = {1\over 1 + \epsilon\  ch (t-\tau)},
\end{equation}

\noindent for any $\tau,$ demonstrating the translational symmetry of the system.  The non-scaled
solution is 

\begin{equation}
\varphi_c = {\sqrt{{2n_3\over J_3}}}\ \ \  {1\over 1 + \epsilon\  ch (t-\tau)}.
\end{equation}

\noindent ${\tilde\varphi}_c$ is shown in Fig.3 as a function of time.  As one can see clearly in
both Figs.2 and 3, $\tilde{\varphi}_c$ starts from 0 in the remote past.  It remains almost zero
for $\tau-t >>1$ because of the exponential behavior of (20).  It increases significantly from zero
in region of $t$ near $\tau.$  At $t = \tau\ \  \tilde{\varphi}_c$ reaches its maximum value
$(1+\epsilon)^{-1}.$  This is also evident in Fig.2, since this value is also the point where the
effective potential vanishes.

\begin{figure}
\begin{center}
\epsfbox[0 0 235 133]{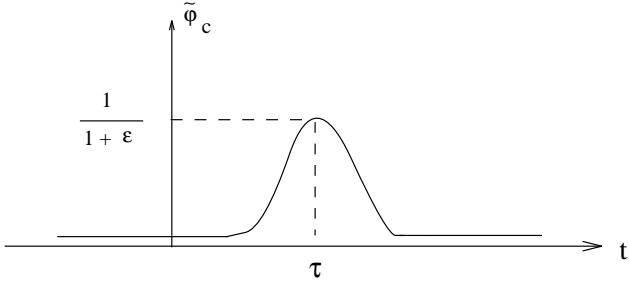}
\caption{The ``bounce'' solution to Eq.(16) for fixed $\epsilon^2 >1.$}
\end{center}
\end{figure}

\indent The calculation of the integrals appearing in (15), as functions of $\epsilon^2,$ is given
in Appendix A.  One obtains

\begin{displaymath}
J_3 = - {3\over (\epsilon^2-1)^2} + {\epsilon^2+2\over (\epsilon^2-1)^{5/2}}
tan^{-1}{\sqrt{\epsilon^2-1}}
\end{displaymath}

\begin{equation}
J_4= {4\epsilon^2+11\over 3(\epsilon^2-1)^3} - {3\epsilon^2+2\over (\epsilon^2-1)^{7/2}}
tan^{-1}{\sqrt{\epsilon^2-1}}
\end{equation}

\begin{displaymath}
\dot{J}_2 = {1\over 2} [J_3 + (\epsilon^2-1) J_4].
\end{displaymath}

\noindent The calculation of the prefactor ${\cal {N}}_\varphi$ is given in Appendix B.  After
isolating the zero mode, it involves the calculation of the determinant of a nontrivial non-local
operator.  We find

$${\cal{N}}_\varphi=\beta{\sqrt{n_3}}\tilde{\cal{N}}_\varphi$$

\begin{equation}
\tilde{\cal{N}}_\varphi ={\sqrt{{8J_4\over \pi[6\dot{J}_2(J_3-J_4) -2\epsilon^2J_3J_4]}}},
\end{equation}

\noindent where $\beta$, coming from the zero mode, is the ``volume''
of the 1-dimensional Euclidean time, and the ${\sqrt{n_3}}$ factor in
(23) comes from an identical factor appearing in the relation (14)
between $\varphi_c$ and ${\tilde\varphi}_c.  \ \tilde{\cal N}_\varphi$
is a function only of $\epsilon^2.$ One can show that it is an
increasing function for $\epsilon^2 > 1.$ The limit

\begin{equation}
\tilde{\cal N }_\varphi^2 
 \mathrel{{{}\atop{\longrightarrow}}\atop{{\epsilon^2\to 1}\atop{}}}15/\pi 
\end{equation}

\noindent shows that our solution is stable, that is it minimizes the effective action.

\section{Asymptotic results for the large $n_3,n_4$ perturbative coefficients and a numerical
application}

\indent Because of the factorial growth of the perturbative coefficients in (10), which originates
from the factor $exp(-S_{eff}[\varphi_c]),$ one can write the leading asymptotic series for Eq.(2)
as

\begin{equation}
E(\alpha) - E(0) = \lim_{\beta \to \infty}\left(-{Z(\alpha)\over \beta}\right).
\end{equation}

\noindent The error from this approximation appears only in the non-leading terms of the series
\cite{Itz}.  Using Eqs.(10) and (23) one sees that the ``volume'' $\beta$ drops out and the ground
state energy is

\begin{equation}
E(\alpha) = {1\over 2} + \sum_{n_3,n_4} \alpha^{n_3+n_4} E(n_3,n_4)
\end{equation}

\noindent the coefficients being

\begin{equation}
E(n_3,n_4) = {(-1)^{n_4+1} 2{\sqrt{n_3}}  
{\tilde{\cal N}}_\varphi\  e^{-S_{eff}[\varphi_c]}\over
(2n_3)! n_4! 2^{n_4}}.
\end{equation}

\noindent Formulas (17), (19), (22), (23), (27) give complete results for the perturbative
coefficients when $n_3$ and/or $n_4$ are large. 

\indent It is not difficult to find
 $E(n_3,n_4)$ numerically.  For given $n_3$ and $n_4$ we first solve Eqs.(17) and (22)
numerically to determine $\epsilon^2.$  Then we get $J_3,J_4,\dot{J}_2$ from (22),
$S_{eff}[\varphi_c]$ from (19), ${\tilde{N}}_\varphi$ from (23) and finally $E(n_3,n_4)$ from
(27).  As an application we fix the order to be $n=n_3+n_4=10.$  We find the values $E(n_3,n_4)$
presented in Table 1.  Because of the alternating sign of the series, delicate cancellations
occur when we sum the terms and the result is

\begin{equation}
\sum_{n_3=0}^{10}E(n_3,10-n_3) =\ - 2.019 \times 10^{11}
\end{equation}

\noindent which is 4 orders of magnitude smaller than the largest term.  It is a well-known
result\cite{Bre}, which we also derive using a different approach later in this paper, that

\begin{equation}
E(\alpha) = \sum_nE_n\alpha^n = \sum_n {-3^{n+1}n!\over \pi}\ \alpha^n
\end{equation}

\noindent neglecting preasymptotics corrections which behave like $1/n$ compared to
$E_n$\cite{Bre,Fal}.  From (29) we obtain

\begin{table}
\begin{center}
\epsfbox[0 0 261 218]{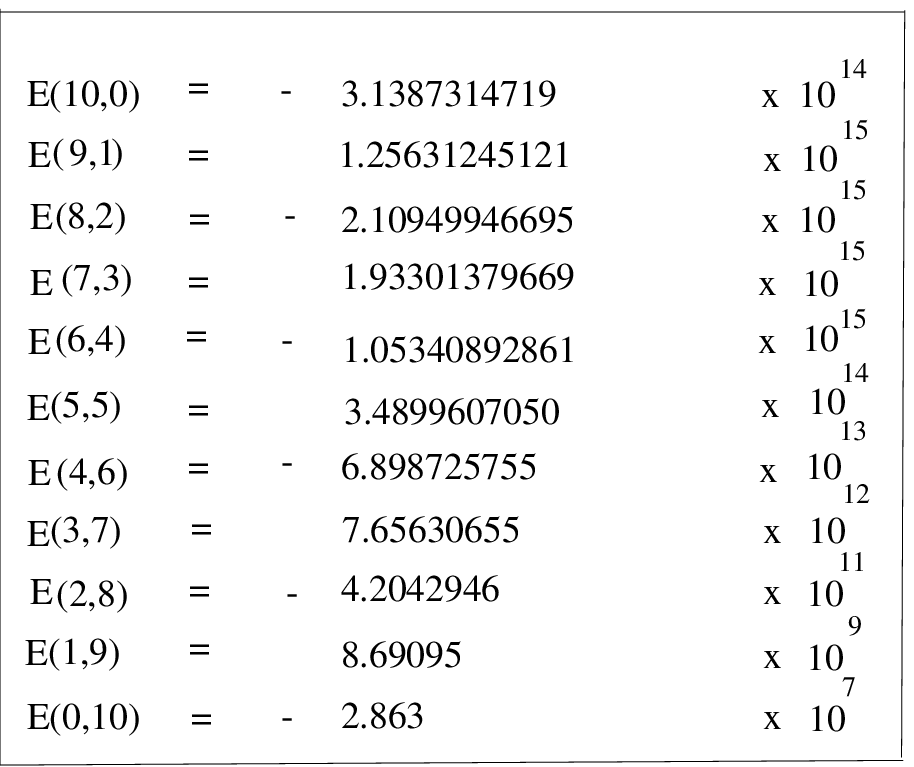}
\caption{$E(n_3,n_4)$ for $n_3 + n_4 = n = 10.$}
\end{center}
\end{table}


\begin{equation}
E_{10} = - 2.046 \times 10^{11}
\end{equation}

\noindent Comparing (28) and (30) we see that the error is 1.31\%.  As\ $n$\ becomes larger, the
error becomes smaller, and also both (27) and (29) are getting closer to the exact value.  Recall
that our approach is to use the steepest descent method around the saddle point $\varphi_c.$  We
calculate only the Gaussian fluctuations.  This will lead to equation (29) as we show in the
following sections.

\section{Borel transformation of the series}

\indent Neglecting the, unimportant for our purposes, zeroth order term $E(0) = 1/2$ we can write
(26) as

\begin{equation}
E(\alpha) = {1\over \alpha} \int db\ e^{-{b\over \alpha}} \sum_{n_3,n_4} {E(n_3,n_4)
b^{n_3+n_4}\over (n_3+n_4)!},
\end{equation}

\noindent or

\begin{equation}
E(\alpha) = {1\over \alpha} \int db\  e^{-{b\over \alpha}}\tilde{E}(b)
\end{equation}

\noindent where the Borel transform $\tilde{E}(b)$ of the ground state energy is given by the
series

\begin{equation}
\tilde{E}(b) = \sum_{n_3,n_4} \ {b^{n_3+n_4}(-1)^{n_4+1}\ 2{\sqrt{n_3}}\ \tilde{\cal N}_\varphi
e^{-S_{eff}[\varphi_c]}\over (2n_3)!n_4!(n_3+n_4)! 2^{n_4}}.
\end{equation}

\noindent In order to show that (27) is in exact agreement with (29) we shall show that
$\tilde{E}(b),$ as given by (33), leads to a simple pole in the Borel plane with the residue
demanded by (29).  This requires transforming the $n_4$-sum in (33) into an integral and summing
over $n_3,$ tasks to which we now turn.

\section{Analytical continuation in the complex $n_4$ plane and summation over the 3-vertices}

\indent The summand in (33) has a very complicated dependence on $\epsilon^2$ and the terms are of
alternating sign in $n_4.$  The summation over $n_4$ cannot be done directly.  We replace this
summation by an integration over a contour in the complex $n_4$ plane in the following way.  In
general, let

\begin{equation}
\sigma = \sum_{n_4=0}^\infty (-1)^{n_4+1}\ f(n_4)
\end{equation}

\noindent then

\begin{equation}
\sigma = {i\over 2} \int_{c^\prime} {dn_4f(n_4)\over sin  \pi  n_4},
\end{equation}

\noindent where the contour $c^\prime$ includes all non-negative integers and is shown in Fig.4. 
Now let $n_4= - \tilde{n}_4,$ then

\begin{equation}
\sigma = {i\over 2} \int_c {
d\tilde{n}_4 f(-\tilde{n}_4)\over sin  \pi  \tilde{n}_4}
\end{equation}

\noindent where $c$ is also shown in Fig.4.  Using

\begin{equation}
\tilde{n}_4!(-\tilde{n}_4)! = {\pi\tilde{n}_4\over sin  \pi  \tilde{n}_4}
\end{equation}

\noindent we finally get

\begin{equation}
\sigma = {i\over 2\pi} \int_c{d\tilde{n}_4
\over \tilde{n}_4} f(-\tilde{n}_4)\tilde{n}_4! (-\tilde{n}_4) !.
\end{equation}

\begin{figure}
\begin{center}
\epsfbox[0 0 315 104]{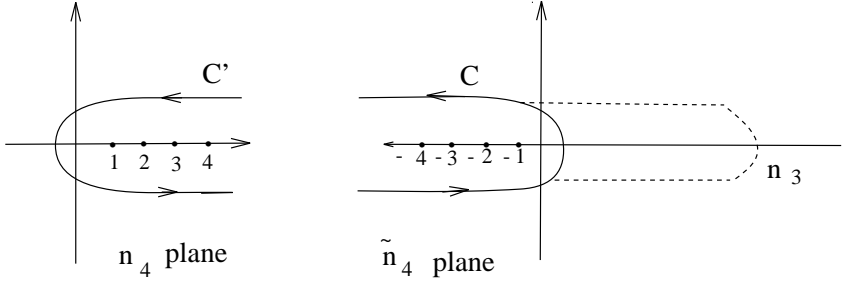}
\caption{Contours of integration in the complex plane.}
\end{center}
\end{figure}

\noindent Therefore one can write $\tilde{E}(b)$ as

\begin{equation}
\tilde{E}(b) = {i\over \pi} \sum_{n_3}\int_c d\tilde{n}_4
{b^{n_3-\tilde{n}_4}2^{\tilde{n}_4}\tilde{n}_4!{\sqrt{n_3}}\tilde{\cal N}_\varphi
e^{-S_{eff}[\varphi_c]}\over \tilde{n}_4(2n_3)! (n_3-\tilde{n}_4)!}.
\end{equation}

\noindent Notice, as is shown in Fig.4, that one can extend the contour  $c$ in (39) to include
positive integers too, so long as $\tilde{n}_4 < n_3.$   This will be important since, as we shall
see, the dominant contribution will come from large and positive $\tilde{n}_4$ close to $n_3.$

\indent In equation (39) one has to let $n_4 \to - \tilde{n}_4$ in $S_{eff}[\varphi_c]$ and
analytically continue the integrals $J_3,J_4, \dot{J}_2$ for complex $\epsilon^2$  as given in
Appendix A.  The variable $\epsilon^2$ clearly becomes

\begin{equation}
\epsilon^2 = 1 - {\tilde{n}_4\over n_3}\ {J_3\over J_4}.
\end{equation}

\noindent Using Stirling's approximation for the factorials, (39) becomes

\begin{equation}
\tilde{E}(b) = {i\over {\sqrt{4\pi^3}}} \sum_{n_3} \int_c d\tilde{n}_4\ {\tilde{\cal N}_\varphi
e^{-n_3A}\over {\sqrt{\tilde{n}_4(n_3-\tilde{n}_4)}}}
\end{equation}

\noindent where\ A\ is a function of $\epsilon^2$ and \ b\  and is given by

\begin{equation}
A = {2\dot{J}_2\over J_3} log {\dot{J}_2\over b}\ -\ {(1-\epsilon^2) J_4\over J_3}\ log
(1-\epsilon^2).
\end{equation}

\indent Since \ A\ depends on $n_3,\tilde{n}_4$ only through
their ratio, we change variables from $n_3$ and $\tilde{n}_4$ to $n_3$ and $r$ where

\begin{equation}
\tilde{n}_4=n_3r
\end{equation}

\noindent and hence

\begin{equation}
d\tilde{n}_4= n_3 dr.
\end{equation}

\noindent Then Eq.(41) becomes

\begin{equation}
\tilde{E}(b) = {i\over {\sqrt{4\pi^3}}} \int {dr \tilde{\cal N}_\varphi\over {\sqrt{r(1-r)}}}
\sum_{n_3=0}^\infty e^{-n_3A}.
\end{equation}

\noindent The summation over the 3-vertices is now trivial leading to

\begin{equation}
\tilde{E}(b) = {i\over {\sqrt{4\pi^3}}} \int {dr \tilde{\cal{N}}_\varphi\over
{\sqrt{r(1-r)}}(1-e^{-A})}.
\end{equation}

\section{The singularity in the Borel plane}

As is clear from (46) the singularity of $\tilde{E}(b)$ will come at a value of b where\ A\
vanishes.  It is not too difficult to see that the important region of integration in equation
(46) is for small $\epsilon^2.$  Defining

\begin{equation}
T = - log {\epsilon^2\over 4}
\end{equation}

\noindent and using the expansions of Appendix A we have

\begin{equation}
\tilde{\cal N}_\varphi = {\sqrt{6 T\over \pi}}
\end{equation}

\begin{equation}
r = 1-{2\over 3T}
\end{equation}

\begin{equation}
dr ={2\over 3T^2} dT
\end{equation}

\noindent
and

\begin{equation}
A = {2\over T} (\delta b - 2e^{-T}),
\end{equation}

\noindent where $\delta b = {1/3} - b$ is small.  Substitution of the above in (46) gives

\begin{equation}
\tilde{E}(b) = {i\over 2\pi^2} \int {dT\over \delta b - 2e^{-T}}.
\end{equation}

\begin{figure}
\begin{center}
\epsfbox[0 0 114 111]{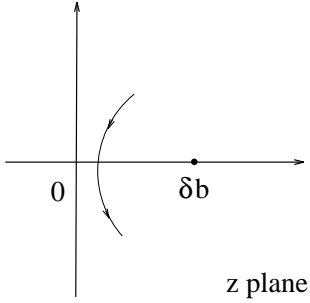}
\caption{The contour in the \ z\ plane near the pole.}
\end{center}
\end{figure}

\noindent Defining $z = 2 \ e^{-T}$\  and using\  $dz = - z dT$  we get

\begin{equation}
\tilde{E}(b) = {i\over 2\pi^2}\int {dz\over z(z-\delta b)}.
\end{equation}

\noindent where the integration contour in (53) is shown in Fig.5, in the region where a pinch
occurs.  The integrand has a first-order pole at $\delta b$ and the residue is $1/\delta b.$ 
Thus we finally find

\begin{equation}
\tilde{E}(b) = - {1\over \pi} \ \ {1\over {1\over 3}-b}
\end{equation}

\noindent which is the well-known result\cite{Bre,Zub,Bog}.  The Borel transform of the ground
state energy has a 1st order pole at $b=1/3$ with residue $1/\pi.$  Using (54), Eq.(32) gives the
leading asymptotic series in Eq.(29).  

\indent Finally we note that as $1/3-b$ becomes small the pinch in the z-integration in (53) at
$z=0$ (or at $T=\infty$) corresponds to a pinch of  the $r-$integration contour in (46) at $r=1$
as is evident from (49).  This means that one can view the pole in $\tilde{E}(b)$ at $b=1/3$ as
being determined by a particular ``graph'' having $n_3 \approx - n_4 = \tilde{n}_4$ large, that is
a large but negative number of 4-vertices.

\section{The $I-\bar{I}$ pair and comments}

\indent The classical solution (21) can also be expressed as

\begin{equation}
\varphi_c={\sqrt{{n\over \dot{J}_2}}}\  \ {1\over 1+\epsilon ch (t-\tau)}
\end{equation}

\noindent where $n=n_3+n_4=n_3-\tilde{n_4}.$  Therefore $\varphi_c$ grows with the square root
of the order in the perturbation series\cite{Itz}.  When $\epsilon^2$ is small, $\dot{J}_2$ is
close to $1/3$ leading to

\begin{equation}
\varphi_c \approx {\sqrt{3n}}\ {1\over 1 + \epsilon ch (t-\tau)}.
\end{equation}

\noindent Now consider a ``standard'' approximate solution \cite{Bre,Zub,Bog} to the equation of
motion

\begin{equation}
-\ddot{\varphi} + \varphi = 3{\sqrt{\alpha}}\ \varphi^2 - 2\alpha\varphi^3
\end{equation}

\noindent corresponding to the Lagrangian given by (1)

\begin{equation}
\varphi = {1\over{\sqrt{\alpha}}} \{{1\over e^{-(t-\tau + {\tilde{T}\over 2})}+1}\ +\ {1\over
e^{t-\tau-{\tilde{T}\over 2}}+1}-1\}.
\end{equation}

\noindent The instanton has its center at $\tau -\tilde{T}/2$ and the antiinstanton at $\tau +
\tilde{T}/2,$ their separation being $\tilde{T}.$  When $\tilde{T}$ is large, (58) is an
approximate solution to (57).  It may be written in the form

\begin{equation}
\varphi = {1\over {\sqrt{\alpha}}}{th {\tilde{T}\over 2}\over 1 + (ch{\tilde{T}\over 2})^{-1}ch(t-\tau)},
\end{equation}

\noindent and as $\tilde{T} \to \infty,$ or $\tilde{\epsilon}^2 = 4 e^{-\tilde{T}}\to 0,$  one
finds

\begin{equation}
\varphi \approx {1\over {\sqrt{\alpha}}}\ {1\over 1 + \tilde{\epsilon} ch (t-\tau)}.
\end{equation}

\begin{figure}
\begin{center}
\epsfbox[0 0 339 126]{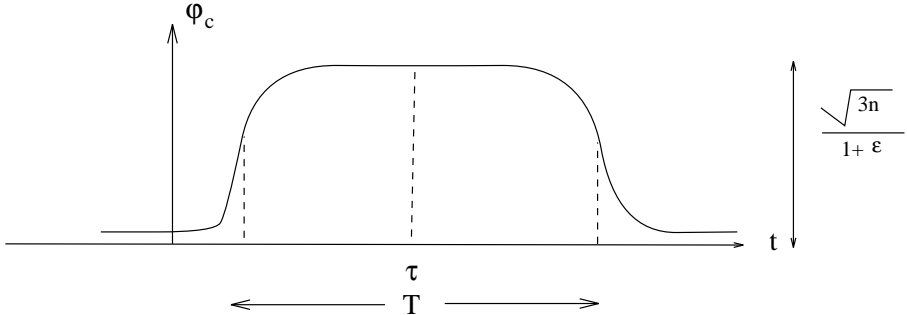}
\caption{The $I-\bar{I}$ pair.}
\end{center}
\end{figure}

\noindent Comparing (56), (60) we see that our exact solution of $S_{eff}$ represents an
$I-\bar{I}$ pair centered at $t=\tau$ and the separation is $T = - \log (\epsilon^2/4).$ 
This is shown in Fig.6.  The small $\epsilon^2$ term in (16) is the difference between equations
(16) and (57), but this is what allows us to have an exact $I-\bar{I}$ solution.  Fig.7 shows
the effective potential for fixed small $\epsilon^2.$  The conclusion is that the solution of the
differential equation (16) which describes large order graphs is an $I-\bar{I}$ pair.  This pair
is a large fluctuation of the vacuum, but it has zero topological charge, and it emerges in a
perturbative approach.

\begin{figure}
\begin{center}
\epsfbox[0 0 190 115]{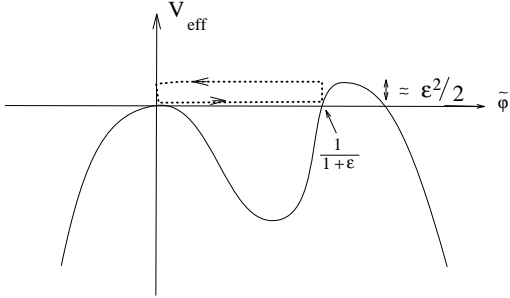}
\caption{The effective potential for small $\epsilon^2.$ The dotted line represents the
classical path.}
\end{center}
\end{figure}

\indent It is also instructive to compare the coefficients ${\sqrt{3n}}$ and
$1/{\sqrt{\alpha}}$ in (56), (60).  These coefficients become equal when

\begin{equation}
3\alpha n = 1.
\end{equation}

\noindent To interpret this we differentiate $E_n$ of Eq.(29) to see when the asymptotic series
starts to diverge.  One finds that ${dE_n\over dn} = 0$ exactly when (61) is satisfied so that the
solution, (56), determining the large orders of the perturbation theory agrees with the
$I-\bar{I}$ approximate solution of the original Lagrangian just when the perturbation series
starts to diverge. 

\indent Finally, we make a couple of comments.  As mentioned earlier the value of the effective
action does not depend on the position $\tau$ of the center of the $I-\bar{I}$ pair.  As can be
seen from Appendix  B, the existence of this ``collective coordinate'' results in a factor

\begin{equation}
\Delta = {\sqrt{{2n_3\dot{J}_2\over J_3}}} = {\sqrt{n}}.
\end{equation}

\noindent This is equivalent to the statement that in a general interacting theory with coupling
$\alpha,$ each collective coordinate, which comes from a symmetry, results in a factor
$1/{\sqrt{\alpha}}.$  One should also comment on the quasizero mode\cite{Yun} which normally
appears as a fluctuation of the $I-\bar{I}$ separation.  In Appendix  B  we calculate the
determinantal ratio $det^\prime(K+U)/det K_0$ which is $(18T^2)^{-1}$ for
large 
$T.$  It looks as if there exists a small eigenvalue with eigenfunction, say, $\eta_1.$  However,
the fluctuation of the action along $\eta_1$ is large.  From the scaling relation (14) one has

\begin{equation}
\eta_1 \sim {\sqrt{{n_3\over J_3}}} \sim {\sqrt{{n_3\over T}}}.
\end{equation}

\noindent Therefore, there is an additional overall factor of $n_3/T$ coming into the effective
action (see (B2) and (B3)) leading to

\begin{equation}
\Delta S_{eff} \sim \lambda_1 \eta_1^2 \sim n_3/T^3
\end{equation}

\noindent from the eigenmode corresponding to $\eta_1.$  From (45), (46), and (51) we see that the
important region of $n_3$ is

\begin{equation}
n_3 \sim {1\over A}\sim {T\over \epsilon^2}.
\end{equation}

\noindent Thus (64) becomes

$$\Delta S_{eff} \sim {1\over \epsilon^2
T^2} \mathrel{{{}\atop{\longrightarrow}}\atop{\epsilon^2\to 0}}\infty.\eqno(66)$$

\vskip 20pt

\noindent{\bf APPENDIX A}
\vskip 10pt
\indent Here we calculate the integrals $J_3, J_4$ and $\dot{J}_2$ as functions of $\epsilon^2,$
first for $\epsilon^2 > 1.$  We start from the integral

$$J(a) = \int_{-\infty}^\infty {dt\over a + \epsilon ch t} = {2\over {\sqrt{\epsilon^2- a^2}}}
tan^{-1} {\sqrt{{\epsilon^2\over a^2}-1}}\eqno(A1)$$

\noindent for $a < \epsilon.$  Differentiating with respect to\  $a\ $ and setting $a=1$ we get
$J_3,J_4$ from

$$J_{n+1}= \int_{-\infty}^\infty {dt\over (a + \epsilon ch t)^{n+1}} = {(-1)^n\over n!}\
{d^nJ(a)\over da^n}\bigg\vert_{a=1}.\eqno(A2)$$

\noindent Thus

$$J_3 = -\ {3\over (\epsilon^2-1)^2}\ +\ {\epsilon^2+ 2\over (\epsilon^2-1)^{5/2}}\ tan^{-1}
{\sqrt{\epsilon^2-1}},\eqno(A3)$$

\noindent and

$$J_4 = {4\epsilon^2+11\over 3(\epsilon^2-1)^3}\ -\ {3\epsilon^2+2\over (\epsilon^2-1)^{7/2}}
tan^{-1}{\sqrt{\epsilon^2-1}}.\eqno(A4)$$

\noindent One can obtain  $\dot{J}_2$ by using the differential equation (16):

$$-\ddot{\tilde{\varphi_c}} + \tilde{\varphi_c} = 3\tilde{\varphi_c}^2
+ 2(\epsilon^2-1)\tilde{\varphi_c}^3.\eqno(A5)$$

\noindent Multiplying (A5) by $\dot{\tilde{\varphi_c}}$ and integrating one obtains

$$\dot{J}_2 - J_2 = - 2J_3 - (\epsilon^2-1) J_4.\eqno(A6)$$

\noindent Multiplying (A5) by $\tilde{\varphi_c}$ and integrating gives

$$\dot{J}_2 + J_2 = 3J_3 + 2(\epsilon^2-1) J_4.\eqno(A7)$$

\noindent Equations (A6) and (A7) give

$$\dot{J}_2 = {1\over 2} [J_3 + (\epsilon^2-1) J_4].\eqno(A8)$$

\indent One may use the above equations for complex $\epsilon^2.$  Since we need the integrals
for small and complex $\epsilon^2$ we write them in the more convenient form

$$J_3 = - {3\over (1-\epsilon^2)^2} + {2 + \epsilon^2\over (1-\epsilon^2)^{5/2}}
th^{-1}{\sqrt{1-\epsilon^2}}\eqno(A9)$$

$$J_4 = - {11 + 4\epsilon^2\over 3(1-\epsilon^2)^3} + {2+3\epsilon^2\over (1-\epsilon^2)^{7/2}}
th^{-1}{\sqrt{1-\epsilon^2}},\eqno(A10)$$

\noindent where $\dot{J}_2$ is still given by (A8).  The expansions to order $\epsilon^2,$ for
small $\epsilon^2$ or equivalently large $T = - log (\epsilon^2/4),$ are:

$$J_3 = T-3 + (12T -26)\  {\epsilon^2\over 4}\eqno(A11)$$

$$J_4 = T-{11\over 3} + (20T - {154\over 3})\ {\epsilon^2\over 4}\eqno(A12)$$

$$\dot{J}_2={1\over 3} - (2T - {16\over 3})\ {\epsilon^2\over 4}\eqno(A13)$$

$${\tilde{n}_4\over n_3} = 1 - {2\over 3(T-3)} + {4[T(3T-11) + 11]\over 3(T-3)^2}\
{\epsilon^2\over 4}\eqno(A14)$$

\noindent where we have used (40) to derive (A14).
\vskip 20pt
\noindent {\bf Appendix B}
\vskip10pt
\noindent Here we calculate the prefactor ${\cal N}_\varphi.$ The effective action is given by
(9).  Its second derivate evaluated at the saddle point $\varphi_c$ is given by

$${\delta^2S_{eff}\over \delta\varphi(t^\prime)\delta\varphi(t)}\bigg\vert_{\varphi_c} =
\{{-d^2\over dt^2}\ + 1-6\tilde{\varphi}_c(t-\tau)
-6(\epsilon^2-1)\tilde{\varphi}_c^2(t-\tau)\}\delta(t-t^\prime)$$

$$+ {9\over J_3} \tilde{\varphi}_c^2(t-\tau)\tilde{\varphi}_c^2(t^\prime-\tau) +
{8(\epsilon^2-1)\over J_4}
\tilde{\varphi}_c^3(t-\tau)\tilde{\varphi}_c^3(t^\prime-\tau).\eqno(B1)$$

\noindent Let $\eta(t)$ be the small fluctuation around $\varphi_c(t-\tau).$  Then the small
fluctuation action is

$$\Delta S_{eff} = {1\over 2} \int dt dt^\prime {\delta^2S_{eff}\over
\delta\varphi(t^\prime)\delta\varphi(t)}\bigg\vert_{\varphi_c}\eta(t)\eta(t^\prime).\eqno(B2)$$

\noindent Substituting (B1) into (B2) we can write $\Delta S_{eff}$ in a convenient bracket
notation\cite{Itz} as

$$\Delta S_{eff} = {1\over 2} < \eta\vert (K +
U)_\tau\vert\eta>\eqno(B3)$$

\noindent with

$$K=K_0 - 6\tilde{\varphi}_c-6(\epsilon^2-1)\tilde{\varphi}_c^2\eqno(B4)$$

$$U={9\over J_3} \vert\tilde{\varphi}_c^2 ><\tilde{\varphi}_c^2\vert + {8(\epsilon^2-1)\over J_4}
\vert\tilde{\varphi}_c^3 >< \tilde{\varphi}_c^3\vert.\eqno(B5)$$

\noindent The index $\tau$ means that the argument of $\tilde{\varphi}_c$ is $t-\tau.$

\indent Thus the prefactor is

$${\cal{N}}_\varphi = {\int\ D\eta\  exp\{-{1\over 2} <\eta\vert(K + U)_\tau\vert\eta >\}\over
\int\  D\eta\  exp\{-{1\over 2} <\eta\vert K_0\vert\eta>\}}.\eqno(B6)$$

\noindent The nonlocal operator $(K+U)_\tau$ has a zero mode given by $\dot{\tilde
\varphi}_c(t-\tau).$  First notice that

$$U\vert \dot{\tilde \varphi}_c > = 0\eqno(B7)$$

\noindent since $< \tilde{\varphi}_c^2\vert \dot{\tilde \varphi}_c > \ =\ <\tilde{\varphi}_c^3\vert
\dot{\tilde \varphi}_c > = 0$ because $\tilde{\varphi}_c$ is even while $\dot{\tilde \varphi}_c$
is odd.  Also

$$K\dot{\tilde \varphi}_c = K_0\dot{\tilde \varphi}_c - 6\tilde{\varphi}_c\dot{\tilde
\varphi}_c - 6(\epsilon^2-1) \tilde{\varphi}_c^2 \dot{\tilde \varphi}_c$$

$$={d\over dt} (3\tilde{\varphi}_c^2 +
2(\epsilon^2-1)\tilde{\varphi}_c^3) - 6\tilde{\varphi}_c\dot{\tilde \varphi}_c -
6(\epsilon^2-1)\tilde{\varphi_c}^2\dot{\tilde \varphi}_c$$

\noindent implies

$$K\dot{\tilde \varphi}_c = 0.\eqno(B8)$$

\noindent Therefore

$$(K+U)_\tau \vert\dot{\tilde \varphi}_c(t-\tau) > = 0.\eqno(B9)$$

\noindent Since $<\dot{\tilde \varphi}_c \vert \dot{\tilde\varphi_c}> = \dot{J}_2$  we define the
zero frequency normalized eigenfunction by

$$\eta_0(t-\tau) = {\dot{\tilde \varphi}_c(t-\tau)\over {\sqrt{\dot{J}_2}}}.\eqno(B10)$$

\indent Now use the Fadeev-Popov trick to change the zero eigenvalue into a collective coordinate
integration

$$1=\int d\tau \delta (<\eta(t)\vert\eta_0(t-\tau)>) \Delta\eqno(B11)$$

\noindent where

$$\Delta = {d\over d\tau} < \eta(t)\vert\eta_0(t-\tau)>.\eqno(B12)$$

\noindent Equation (B12) leads to

$$\Delta = < \dot{\varphi}(t)\vert\eta_0(t-\tau)>.\eqno(B13)$$

\noindent In the sense of the steepest descent method $\dot{\varphi}(t)$ may be approximated by
$\dot{\varphi}_c(t-\tau)$ so that

$$\Delta = <\dot{\varphi}_c(t-\tau)\vert\eta_0(t-\tau)>\eqno(B14)$$

\noindent giving finally

$$\Delta = {\sqrt{{2n_3\dot{J}_2\over J_3}}}.\eqno(B15)$$

\noindent The identity (B11) is inserted in the functional integral expression (B6) for
${\cal N}_\varphi$ giving

$${\cal N}_\varphi = {\Delta \int d\tau \int D\eta\delta(<\eta\vert\eta_0(t-\tau)>)exp\{-{1\over
2}<\eta\vert(K+U)_\tau\vert\eta>\}\over \int D\eta\  exp\{-{1\over
2}<\eta\vert\ K_0\vert\eta>\}}.\eqno(B16)$$

\noindent In the numerator of (B16) the functional integral is independent of $\tau,$ so we can
choose, say, $\tau=0.$  The integration over $\tau$ is trivial and gives the one-dimensional volume
of time $\beta.$  Thus

$${\cal{N}}_\varphi = {\beta\Delta\over {\sqrt{2\pi}}}{\sqrt{{det K_0\over det^\prime(K+U)}}}\eqno(B17)$$

\noindent where the prime means that the zero eigenvalue is omitted because of the
$\delta-$function in (B16).  

\indent Now we turn to the nonlocal part U.  In general

$$det(K+\sum_{\alpha =1}^N\vert \alpha ><\alpha\vert) = det K \cdot det (1 + \sum_{\alpha = 1}^N
K^{-1}\vert \alpha><\alpha\vert)\eqno(B18)$$

\noindent where $K^{-1}$ is well-defined and is linear.  The second determinant can be shown to
obey

$$det (1+ \sum_{\alpha = 1}^N K^{-1}\vert\alpha ><\alpha\vert) = det (1+\hat{K}^{-1})\eqno(B19)$$

\noindent where on the right-hand side of (B19) $\hat{K}^{-1}$ is a $N\times N$ matrix with elements

$$\hat{K}_{\alpha\alpha^\prime}^{-1} = <\alpha\vert K^{-1}\vert\alpha^\prime>.\eqno(B20)$$

\noindent In our case $K^{-1}$ is well-defined because we consider the primed determinant.  The
operation of $K^{-1}$ on $\vert \tilde{\varphi}_c^2>, \vert \tilde{\varphi}_c^3>$ is also defined
because these states are orthogonal to the zero mode. Thus

$$det^\prime(K+U)= det^\prime K \cdot det(1+\hat{K}^{-1})\eqno(B21)$$

$$det(1+\hat{K}^{-1}) = (1+{9\over J_3}\hat{K}_{22}^{-1})(1+ {8(\epsilon^2-1)\over J_4} \hat{K}_{33}^{-1})
-{72(\epsilon^2-1)\over J_3J_4} (\hat{K}_{23}^{-1})^2.\eqno(B22)$$

\indent It is not difficult to find the three matrix elements.  Start from the operation of $K$
on $\tilde{\varphi}_c:$

$$K \tilde{\varphi}_c = -3 \tilde{\varphi}_c^2 - 4(\epsilon^2-1) \tilde{\varphi}_c^3$$

$$\tilde{\varphi}_c = - 3 K^{-1} \tilde{\varphi}_c^2 - 4(\epsilon^2-1)
K^{-1}\tilde{\varphi}_c^3.\eqno(B23)$$

\noindent Multiplying (B23) by $\tilde{\varphi}_c^2$ and $\tilde{\varphi}_c^3$ in turn and
integrating gives

$$J_3= - 3\hat{K}_{22}^{-1} - 4(\epsilon^2-1) \hat{K}_{23}^{-1}\eqno(B24)$$

\noindent and

$$J_4 = - 3\hat{K}_{23}^{-1}- 4(\epsilon^2-1) \hat{K}_{33}^{-1}.\eqno(B25)$$

\noindent  A third equation may be obtained from the operation of $K$ on $\tilde{\varphi}_c^2$ in
a similar way:

$$K\tilde{\varphi}_c^2 =   -3\tilde{\varphi}_c^2 + 4\tilde{\varphi}_c^3$$

$$\tilde{\varphi}_c^-3K^{-1} \tilde{\varphi}_c^2 + 4K^{-1}\tilde{\varphi}_c^3$$
 
$$ J_4 = -3\hat{K}_{22}^{-1} + 4\hat{K}_{23}^{-1}.\eqno(B26)$$

\noindent Solving (B24), (B25), (B26) we find the matrix elements and (B22) gives

$$det (1+\hat{K}^{-1}) = 2 - {6 \dot{J}_2(J_3-J_4)\over \epsilon^2J_3J_4}.\eqno(B27)$$

\indent Finally, we need to calculate the ratio of determinants $det^\prime K/det K_0.$  We
consider a finite sized ``volume'' $\beta.$  Then  $K$  has no zero eigenvalue, since translation
symmetry is lost, but there is a small eigenvalue $\lambda_0$ which goes to zero as $\beta \to
\infty.$  The determinantal ratio can be obtained just from he knowledge of the classical
solution.  Following references\cite{Col,Kle} one writes

$${det^\prime K\over det K_0}\ = \lim_{\beta \to \infty}\ {\psi(\beta/2)\over
\lambda_0\psi^{(0)}(\beta/2)}\eqno(B28)$$

\noindent where $\psi(t), \psi^{(0)}(t)$ are zero eigenvalue solutions of $K,K_0$ respectively,
which do not vanish at $t = \beta/2,$ but satisfy the boundary conditions

$$\psi(-\beta/2) = 0, \ \ \ \ \ \ \ \ \  \dot{\psi}(-\beta/2) =1,\eqno(B29)$$

\noindent with analogous boundary conditions for $\psi^{(0)}.$  It is trivial to find
$\psi^{(0)}(\beta/2)$ for the free theory:

$$\psi^{(0)}(t) = sh (t + \beta/2)$$
$$\psi^{(0)}(\beta/2) = sh \beta$$
$$\psi^{(0)}(\beta/2)\mathrel{}{{{}\atop{\longrightarrow}}\atop{\beta\to \infty}} e^\beta/2.\eqno(B30)$$

\noindent Now we need to find $\psi(t).$  We already know one solution $\dot{\tilde \varphi}_c$
satisfying $K\dot{\tilde\varphi}_c=0.$  For convenience we normalize it to be $-{\epsilon\over
2}\dot{\tilde \varphi}_c$ and call it $\psi_1.$  We need only the asymptotic behavior

$$\psi_1 \mathrel{}{{{}\atop{\simeq}}\atop{{t\to\pm \infty}}}\pm e^{-\vert t \vert}  .\eqno(B31)$$

\noindent A second independent solution is

$$\psi_2 \sim \psi_1(t) \int^t {dt^\prime\over \psi_1^2(t^\prime)},\eqno(B32)$$

\noindent which leads to

$$\psi_2 \mathrel{}{{{}\atop{\simeq}}\atop{t\to\pm \infty}} e^{\vert t \vert}.\eqno(B33)$$

\noindent As expected $\psi_2$ has the opposite exponential behavior from $\psi_1,$ and $\psi_2$
is even while $\psi_1$ is odd.  One can write the linear combination of $\psi_1,\psi_2$ which
satisfies (B29) as

$$\psi(t) = {1\over 2}\{\psi_1(-\beta/2) \psi_2(t) - \psi_2(-\beta/2)\psi_1(t)\}\eqno(B34)$$

\noindent and its value at $t=\beta/2\  {\rm as}\  \beta \to \infty$ is

$$\psi(\beta/2) \mathrel{{{}\atop{\longrightarrow}}\atop {\beta\to \infty}}-1
.\eqno(B35)$$

\noindent It remains to get the small eigenvalues $\lambda_0.$  To first order in perturbation
theory the solution $\psi_\beta(t)$ that vanishes at $t = \pm \beta/2$ is

$$\psi_\beta(t) = \psi (t) + {\lambda_0\over 2} \int_{-\beta/2}^t
dt^\prime\psi(t^\prime)[\psi_1(t)\psi_2(t^\prime)-\psi_2(t)\psi_1(t^\prime)].\eqno(B36)$$

\noindent It is obvious that $\psi_\beta(-\beta/2) = 0.$  Requiring $\psi_\beta(\beta/2) =0$ gives
$\lambda_0$ as

$$\lambda_0=2[\psi_1(\beta/2) < \psi\vert \psi_2> - \psi_2(\beta/2) < \psi\vert
\psi_1>]^{-1}.\eqno(B37)$$

\noindent Substituting $\psi$ from (B34) in (B37) and using $<\psi_1\vert \psi_2 > = 0$ gives

$$\lambda_0 = 4[-e^{-\beta}< \psi_2\vert \psi_2> + e^\beta <\psi_1 \vert \psi_1 >
]^{-1}.\eqno(B38)$$

\noindent The dominant term in (B38) is the second one which diverges while the first one remains
finite for $\beta \to \infty.$  We finally obtain

$$\lambda_0 = {16 e^{-\beta}\over \epsilon^2\dot{J}_2}\eqno(B39)$$

\noindent which vanishes when $\beta \to \infty$ as claimed.  Hence the determinantal ratio from
(B28), (B30), (B35), (B39) is

$${det^\prime K\over det K_0}\ =\ -{\epsilon^2\dot{J}_2\over 8}.\eqno(B40)$$

\noindent The minus sign is not surprising since $K$ has a negative eigenvalue.  Recall that the
zero mode is odd with one node, therefore there should be an even mode with no nodes and a
smaller, thus negative, eigenvalue.  This minus sign causes no problem in the stability of the
solution since $det (1 + \hat{K}^{-1})$ is negative too leaving the saddle point $\varphi_c$ a minimum
for the action\cite{Itz}.  Putting everything together from equations (B15), (B17), (B21), (B27)
and (B40) the prefactor as a function of $n_3$ and $\epsilon^2$ is

$${\cal{N}}_\varphi = \beta{\sqrt{n_3}}\ \tilde{\cal N}_\varphi$$

$$\tilde{\cal N}_\varphi = {\sqrt{{8J_4\over \pi[6\dot{J}_2(J_3-J_4)
-2\epsilon^2J_3J_4]}}}.\eqno(B41)$$

\end{document}